# Double photoionization of propylene oxide: a coincidence study of the ejection of a pair of valence-shell electrons

*Stefano Falcinelli,[1,*] Franco Vecchiocattivi,[1] Michele Alagia,[2] Luca Schio,[2,3] Robert Richter,[4] Stefano Stranges,[2,5] Daniele Catone,[6] Manuela S. Arruda,[7,8] Luiz A. V. Mendes,[9] Federico Palazzetti,[7,10] Vincenzo Aquilanti,[6,7] and Fernando Pirani[7,11]*

[1]Dipartimento di Ingegneria Civile ed Ambientale, Università di Perugia, Via G. Duranti 93, 06125 Perugia, Italy.

[2]IOM-CNR Tasc, Km 163.5, Area Science Park, 34149, Basovizza, Trieste, Italy.

[3]Dipartimento di Scienze di Base e Applicate per l'Ingegneria (SBAI), Università di Roma Sapienza, Via A. Scarpa 14, 00161 Roma, Italy.

[4]Elettra-Sincrotrone Trieste, Area Science Park, 34149, Basovizza, Trieste, Italy.

[5]Dipartimento di Chimica e Tecnologie del Farmaco, Università di Roma Sapienza, 00185 Roma, Italy.

[6]Istituto di Struttura della Materia (CNR-ISM), Area della Ricerca di Roma Tor Vergata, Via del Fosso del Cavaliere, 100-00133 Roma, Italy.

[7]Dipartimento di Chimica, Biologia e Biotecnologie, Università di Perugia, Via Elce di Sotto 8, 06123 Perugia, Italy.

[8]Centro de Ciêcias Exatas e Tecnológicas, Universidade Federal do Recôncavo da Bahia, Rua Rui Barbosa 710, 44380-000 Cruz das Almas, BA, Brazil.

[9]Instituto de Fìsica, Universidade Federal da Bahia, Campus Universitario de Ondina, 40210-340 Salvador, BA, Brazil.

[10]Scuola Normale Superiore di Pisa, Piazza dei Cavalieri 7, Pisa I-56126, Italy.

[11]CNR NANOTEC, via G. Amendola 122/D, 70126 Bari, Italy.

[*]Corresponding author: Stefano Falcinelli, Dipartimento di Ingegneria Civile ed Ambientale, Università di Perugia, 06125 Perugia, Italy tel. +39 075 5853862, fax +39 075 5853864, email: stefano.falcinelli@unipg.it

**ABSTRACT**

Propylene oxide, a favorite target of experimental and theoretical studies of circular dichroism, was recently discovered in interstellar space, further amplifying the attention to its role in the current debate on protobiological homochirality. In the present work, a photoelectron-photoion-photoion coincidence technique, using an ion-imaging detector and tunable synchrotron radiation in the 18.0-37.0 eV energy range, permits: (i)-to observe six double-ionization fragmentation channels, their relative yields being accounted for about two-thirds by the couple ($C_2H_4^+$, $CH_2O^+$), one-fifth by ($C_2H_3^+$, $CH_3O^+$); (ii)-to measure thresholds for their openings as a function of photon energy; (iii)-to unravel a pronounced bimodality for a kinetic-energy-released distribution, fingerprint of competitive non-adiabatic mechanisms.

**INTRODUCTION**

Molecular chirality, involving left-right dissymmetry, plays a fundamental role in various areas of life science, at both macroscopic and microscopic scales. Homochirality in terrestrial life[1-3] and the high enantio-selectivity in processes involving biologically important chiral molecules are among the most intriguing aspects in natural phenomena and related chiral technologies. Investigation of molecular enantiomeric nature has therefore a strong impact in various areas of chemistry, such as heterogeneous enantioselective catalysis, photochemical asymmetric synthesis, drug activity, enzymatic catalysis, and chiral surface science involving supramolecular assemblies.

Extensive experiments are currently carried out world-wide in laboratories and at large-scale facilities using circularly polarized light, and detecting chirality in molecules by imaging photoelectron circular dichroism (see for example Ref. 4, and references therein). Since molecular alignment has been found to be of general relevance to control the stereodynamics of elementary chemical processes occurring in the gas phase and at surfaces,[5-8] its role in chirality issues is arguably crucial. Specifically, previous studies[9-11] indicate that molecular directionality and alignment should strongly amplify dichroism and provide stereodynamical mechanisms for discrimination of enantiomers other than *via* circularly polarized light. In the present work, we take initial steps in both these directions: Propylene oxide[12,13] is a prototypical experimentally available chiral molecule, and unique in having been characterized in previous experiments for directional and aligning properties by electrostatic hexapole techniques.[14,15] Such a study is being extended in our laboratory, coupling the mechanical molecular velocity selector method (permitting to control of the velocity dependence of molecular alignment)[5-8] with electron-ion imaging detection using tunable synchrotron radiation with high intensity.[16,17] Previous studies on propylene oxide, performed by exploiting Auger spectroscopy, indirectly estimated the first double ionization threshold,[18,19] allowing the observation of a weak circular dichroism effect[19] after the pionieristic work performed by Prümper *et al.*.[20]

In addition, it has to be noted that the use of propylene oxide, as a prototype chiral molecule in order to investigate the possible $(C_3H_6O)^{2+}$ molecular dication formation by VUV double photoionization, is of particular interest because of the recent astronomical detection of such a neutral chiral molecule in absorption toward the Galactic center.[2]

The importance of the double photoionization, and of the elementary processes induced, has been demonstrated by a number of papers, in particular by Eland and coworkers,[21-23] by extending Zare's approach to photodissociation.[24] Recently, various authors[24,25] studied the role of molecular dications in the chemistry of upper planetary atmospheres (see also Refs. 27, 28 and references therein).

In the experiment reported here, using the same technique successfully employed in previous studies (see Ref. 29 and references therein), the double photoionization of propylene oxide molecules has been investigated in order to identify the leading two-body dissociation channels and measure: i) the threshold energy for the formation of different ionic products; ii) the related branching ratios, and iii) the kinetic-energy-released (KER) distribution of fragment ions at different photon energies. This study is relevant both to fill the gap of missing data on the $(C_3H_6O)^{2+}$ dication energetics and its dissociation dynamics, and to provide the information required for further experimental and theoretical investigations of the interaction between chiral molecules and linearly or circularly polarized radiation. Specifically, interaction of circularly polarized synchrotron light with enantiomers of propylene oxide, at different photon energies, is a prerequisite of perspective investigations aimed at discriminating behavior of enantiomers in the angular and energy distribution of fragments and ejected photoelectrons.

**EXPERIMENTAL**

The experiment has been performed at the ELETTRA Laboratory at the CiPo (Circularly Polarized) beamline. A detailed description of the beamline and the end station is given elsewhere.[30,31] Some characteristics of interest for the present experiment have been recently reported in connection with a time-of-flight (TOF) study of product ions in the double photoionization of various molecules.[32-34] Here is an outline of the experimental setup.

The energy selected synchrotron light beam crosses an effusive molecular beam of the neutral precursor and the product ions are then detected in coincidence with photoelectrons. The propylene oxide molecular beam is prepared by effusion from a glass cylinder containing a commercial liquid sample (a racemic mixture with a 99% nominal purity), and supplied by a a stainless steel needle effusive beam source (with a nozzle having a 1.0 mm diameter) taking advantage of its very high vapor pressure (0.59 bar at 20°C). The CiPo beamline uses a Normal Incidence Monochromator (NIM), equipped with two different holographic gratings, allowing to cover the 8 - 37 eV energy range by means of a Gold (2400 l/mm) and an Aluminum (1200 l/mm) coated grating. The resolution in the investigated photon energy range is about 2.0 - 1.5 meV; spurious effects, due to ionization by photons from higher orders of diffraction, are reduced by the use of the NIM geometry. The molecular beam of the neutral precursor molecule and the VUV light beam cross at right angles. The CiPo beamline provides linearly or circularly polarized radiation. In this experiment we employed only the linearly polarized component.

The incident photon flux and the gas pressure are monitored and the ion yields for each investigated channel were divided by the total ion yield in order to obtain the relative branching ratios as a function of the photon energy (see below). To detect the photoions in coincidence with photoelectrons, ejected from the photoionized neutral molecular precursor, we used the electron-ion-ion coincidence technique. The ion extraction and detection system has been described in detail in Ref. 35 and is shown in Fig. 1. It has been especially designed to measure the cation photofragment momentum vectors in many-body dissociation processes.[35] In our experiment the electron signal is simply used as a start pulse for the two ion time-of-flights, and no information about the photoelectron energies can be recorded.

## RESULTS AND DISCUSSION

*Reaction channels*

The mass spectra detected in the 18–37 eV photon energy range indicate that double photoionization of propylene oxide, $C_3H_6O$, produces mainly the two-fragment ion channels indicated below and in Table I.

$$C_3H_6O + h\nu \rightarrow C_2H_4^+ + CH_2O^+ + 2e^- \quad (1)$$

$$C_3H_6O + h\nu \rightarrow CH_2^+ + C_2H_4O^+ + 2e^- \quad (2)$$

$$C_3H_6O + h\nu \rightarrow CH_3^+ + C_2H_3O^+ + 2e^- \quad (3)$$

$$C_3H_6O + h\nu \rightarrow O^+ + C_3H_6^+ + 2e^- \quad (4)$$

$$C_3H_6O + h\nu \rightarrow C_2H_3^+ + CH_3O^+ + 2e^- \quad (5)$$

$$C_3H_6O + h\nu \rightarrow OH^+ + C_3H_5^+ + 2e^- \quad (6)$$

The first valuable information coming from these results is the identification of dissociation channels, their relative abundance, and the threshold energies (see Table I). We attempt their classification in terms of bond cleavage between the heavy atoms taking into account also a possible H atom migration mechanism as previously observed in other systems.[34,36] Anticipating results from data analysis and referring to the abundance in the third column of Table I, the most relevant classification can be suggested. Therefore, in terms of the relative yield of Table I, the following groups are identified:

(1a) The reaction 1 in Table I [$C_2H_4^+$, $CH_2O^+$] accounts for *ca.* 67% of total average yield. The lighter ion $C_2H_4^+$ is the most abundant in the single ionization mass spectrum of propylene oxide published by NIST;[37] the heavier one is also present but with lower abundance as confirmed by previous experimental results by Liu *et al.*[38] and by Gallegos and Kiser;[39]

(1b) The reaction 5 in Table I [$C_2H_3^+$, $CH_3O^+$] *ca.* 20% abundance, this couple of ions is also abundant in the single ionization mass spectrum,[37,38] and differs from the previous one only by H-exchange;

(2a,b) Reactions 2 and 3 in Table I [$CH_2^+$, $C_2H_4O^+$] and [$CH_3^+$, $C_2H_3O^+$] (8% and 5%, respectively) also are related by H-exchange and involve ions present in moderate abundance in the mass spectra.

(3a, b) Reactions 4 and 6 in Table I [$O^+$, $C_3H_6^+$] and [$OH^+$, $C_3H_5^+$] (2% and 0.2%, respectively) are also related by H exchange and involve minor ions in the mass spectra.

For clarity, only a portion of the recorded mass spectrum, together with the relative part of the ion–ion time-of-flight coincidence diagram including only the two main two body dissociation channels, is reported in Fig. 2, for a photon energy of 37.0 eV. Such a Figure is useful to highlight the recorded ion signal due to the double coincidences produced by the dissociation of the intermediate $(C_3H_6O)^{2+}$ molecular dication (see for example the white dashed oval in Fig. 2), with respect to the signal due to single ionization events. It has to be noted that in all our recorded coincidence plots a direct evidence of the formation of a stable $(C_3H_6O)^{2+}$ molecular dication was not found. In fact, such a double ionized species is not visible in the single ion time of flight spectrum either due to the overlap with the singly ionized fragments (see Fig. 2). Furthermore, the possible formation of a metastable $(C_3H_6O)^{2+}$ molecular dication having a relatively long lifetime was not recorded. In fact, in this type of experiments, it is well known that product ions showing a "tail" in the coincidences plot indicate that the fragmentation reaction occurs in a time longer than the characteristic time window of the apparatus (~50 ns).[40] The lack of such a structure in the coincidence plots collected at all investigated photon energies clearly indicates that, in the double photoionization of propylene oxide in the 18.0-37.0 eV energy range, the intermediate $(C_3H_6O)^{2+}$ molecular dication is a short living transient species having a lifetime shorter than 50 ns (for major details see Refs. 28, 29, 32-34). In Fig. 2 some product ions are shown together with some background peaks. The most relevant part of such a diagram allows us to distinguish the ionic products of the main fragmentation channels, triggered by the electrons produced in the double photoionization process and detected in coincidence, which are $C_2H_4^+$ + $CH_2O^+$ and $C_2H_3^+$ + $CH_3O^+$ product ions. In Fig. 2, on the *x* axis, we report a portion of the TOF mass spectrum of all recorded ions showing the peaks related to fragment ions of reactions (1)-(6), and the intensity of each peak in the mass spectrum as a function of the photon energy, normalized for the total ion yield, permits to determine the relative cross section for each fragmentation channel. The second delay time $t_2$ refers to the flight time of the second ion fragment. At each wavelength the total counts of ion pairs (see the white oval in Fig. 2) triggered by the electrons produced in the double photoionization process and detected in coincidence gives the intensity and the cross section of the two-body fragmentation channels.[34] In order to study the threshold and the energy dependence of the two-body dissociative channels (see Table I), we have integrated the density of the ion pair coincidence events, normalized for the total ion yield, as a function of the photon energy. The results are plotted in Fig. 3(a) for all fragmentation channels (see Table I) except the least intense one that is reported in the sixth row of Table I, while in Fig. 3(b) the relative cross sections for the same channels with their relative threshold energies are shown.

*Threshold energies*

For each open dissociation channel, the threshold energy has been determined by using a Wannier function,[41] as we have already done in previous works.[42-44] In such a procedure, an empirical function is used where the threshold energy is an adjustable parameter varied until the best fit of experimental data. Consequently, the errors reflect the scattering of experimental data In particular, in Fig. 3(b) the investigated fragmentation processes are reported with their respective threshold energies and with their averaged intensities, estimated by adding all collected counts for each pair of ion products at all investigated photon energies. The threshold energies for the two-body fragmentation open channels investigated in our experiment, operating at a photon energy between 18.0 and 37.0 eV (see Table I), are in the range of previous experimental estimations from other laboratories by exploiting the Auger spectroscopy. Following the analysis of the electronic structure of propylene oxide, including core-level photoelectron spectroscopy, X-ray absorption at both C and O K-shells, resonant Auger and normal Auger spectroscopy, performed by Piancastelli *et al.*[18], we were able to extract from the published spectra an estimate of about 29.2 eV for the threshold of the double photoionization of propylene oxide. Such a result was confirmed by Alberti *et al.*[19] in C 1s excitation and ionization of propylene oxide studied by means of photoabsorption, photoemission and photoelectron–Auger electron coincidence experiments using linearly and circularly polarized synchrotron radiation. In our experiment we measured directly the first double photoionization energy for the propylene oxide molecule of 28.3±0.1 eV (see reaction 1 of Table I) that is in reasonable agreement with the indirect estimation of 29.2 eV obtained from the work by Piancastelli *et al.*.[18] It has to be noted that in our experiment we were able to obtain for the first time a direct determination of the first double ionization energy in the photoionization of propylene oxide, and also we identified and measured the threshold energies for each opened two body dissociation channel (see data of Table I). This is a basic information that, combined with other experimental findings provided by the same experiment and discussed below, casts light for the first time on the microscopic mechanism promoted by the removal of two valence electrons, and suggests new experimental probes for their full characterization.

*KER distributions*

The values of the kinetic energy of product ions are determined by a simple analysis based on the method suggested by Lavollée[35,45] in which TOF and position on the detector of the ion allow one to extract the complete information about the linear momentum ($p_x$, $p_y$, $p_z$) for each ion without ambiguity, and then its angular and energy distribution. In particular, this full mapping is reached by using a momentum matching analysis for each ion pair in the coincidence spectra measured at all the investigated photon energies (see Fig. 2). In such an analysis we used a momentum matching filter in order to select the true coincidences related to each recorded two-body dissociation channel, coming out from the Coulomb explosion of the intermediate $(C_3H_6O)^{2+}$ molecular dication, applying the condition expressed by the following relation:

$$S \leq \frac{\sqrt{p_{x1,2}^2 + p_{y1,2}^2 + p_{z1,2}^2}}{|\boldsymbol{p}_1| + |\boldsymbol{p}_2|} \qquad (7)$$

where $p_{x1,2}$, $p_{y1,2}$, and $p_{z1,2}$ are the sum of projections of momentum vectors $\boldsymbol{p}_1$ and $\boldsymbol{p}_2$ of the produced fragment ions 1 and 2 in the observed two-body fragmentation reactions. In our analysis, in order to totally subtract the false coincidences, a value of $S \leq 0.1$ was used.

The KER distributions of the two ion products from the main two-body fragmentation reactions (see Table I), with the exception of the less intense one (reaction 6 of Table I) as obtained from the momentum matching analysis of the ion peaks (see equation (7)), and discussion above), have been analyzed as a function of the photon energy. An important result is that the intensity of the KER peaks depends on the photon energy, but their position and shape do not change appreciably. It is well known that all microscopic photochemical processes leading to a two-body fragmentation, are driven by the dynamics on multidimentional potential energy surfaces leading to both ionic and neutral fragment products which involve avoided crossing, saddle points, and various nonadiabatic couplings, determining the opening towards the exit channels. On the basis of the present experimental findings, we can draw some qualitative considerations on the dynamics associated with the ion-ion separation coordinate (see also below). In particular, the observation that the peak position and shape of KER does not change with the photon energy, suggests that the excess of energy (the difference between the applied photon energy and the appearance energy of the considered channel) is distributed either in the internal degrees of freedom of the fragment ions or in the kinetic energy of the ejected electrons.

The peak position for the various channels are reported in Table I, while in Fig. 4 the KER for reaction (2) are shown. It was not possible to determine KER distributions related to product ions of reaction (6) because of the too low intensity of recorded signals for $OH^+ + C_3H_5^+$ coincidences. In particular, their symmetric shape facilitates easy fitting by a simple Gaussian function. This can be considered as a clear indication that each fragmentation channel involves a well-defined single specific region of the multidimensional potential energy surface, associated to the effective intramolecular interaction within the $(C_3H_6O)^{2+}$ dication frame and responsible for the opening of the various two-body fragmentation channels, at all investigated energies. Therefore, for all investigated fragmentation channels the excess of the used photon energy with respect to the double ionization threshold energy should be released as kinetic energy of ejected electrons. The only exception is constituted by the recorded total KER distribution for $CH_3^+ + C_2H_3O^+$ product ions of reaction (3) shown in Fig. 5 for two investigated photon energies. It is evident that such total KER distributions are characterized by a bimodal behavior, previously observed also in the fragmentation of simpler dications as $CO_2^{2+}$.[46] In the Figure, the data are best-fitted (full line) by the combination of two Gaussian functions (dashed lines), clearly indicating a bimodal behavior, depending on two possible microscopic mechanisms for the two-body fragmentation of $(C_3H_6O)^{2+}$ producing $CH_3^+$ + $C_2H_3O^+$. This is an indication that reaction (3) may occur by two different pathways: in one case (the most important one) the two microscopic mechanisms can involve a direct fragmentation of the $(C_3H_6O)^{2+}$ into $CH_3^+$ + $C_2H_3O^+$ products and different internal energy states of $C_2H_3O^+$; in the second case, the fragmentation of the $(C_3H_6O)^{2+}$ can involve also an internal rearrangement of the molecular bonds of the dication and possible atom migration. Further, it has to be noted that $CH_3^+$ and $C_2H_3O^+$ ion fragments can be also formed in an electronically excited state. Therefore, two distinct mechanisms occur for the dynamics along paths in multidimensional potential energy surfaces of the intermediate ion $(C_3H_6O)^{2+}$ produced by ejection of two electrons, accompanied by its breaking in two fragments. Moreover, from the measured KER's, and assuming a simple Coulomb repulsion as the main responsible of the fragmentation, it is possible to evaluate the distances where ions start separating. Estimated values of 2.4 and 3.6 Å represent for the main and less important channel, respectively, the separation distance where a transition, between the internal molecular rearrangement to the opening of the Coulomb repulsion channel, occurs. For the other channels the same KER analysis suggests that only one critical distance is operative, intermediate between two values given above.

Experiments are planned at the Elettra Synchrotron Facility with the aim to measure the anisotropy parameter[24,34] in the angular distribution of dissociation ion products as a function of the photon energy, in order to provide insight on the dynamics of the dication fragmentation.

The observed prominent bimodality (Fig. 5) is interestingly reminiscent of recent much investigated phenomena in photodissociation of molecules in neutral fragments. In the simplest ester $HCOOCH_3$ dissociating in $CH_3OH$ and $CO$,[47,48] a fraction of the latter is released later than expected: the simplified picture that may be adopted here is the promotion of the intermediate state to an upper slope of a conical intersection, followed either by a nearly vertical decay close to transition states (the direct mechanism leading to faster products) or delayed decay due to partial trapping in the conical intersection (the "roaming" mechanism leading to slower products).

**SUMMARY AND CONCLUDING REMARKS**

In this paper, we presented a study of the double photoionization of a prototypical chiral molecule, propylene oxide, promoted by direct ejection of two valence electrons, as a first step of an experimental investigation able to highlight possible differences, characterizing the interaction of polarized light with chiral systems, such as the angular distribution of photo-emitted electrons and of produced ions. This study has been performed by using linearly polarized synchrotron radiation in order to identify the leading two-body dissociation channels and to measure: (i) the threshold energy for the different ionic products formation; (ii) the related branching ratios, and (iii) the kinetic-energy-released distribution of fragment ions at different photon energies. This exploratory study provides previously unavailable data on $(C_3H_6O)^{2+}$ dication energetics and molecular dissociation dynamics, that represent mandatory information for further experimental and theoretical investigations of the interaction between chiral molecules and linearly or circularly polarized light.

In our experiment we were able to measure directly the first double ionization energy in the photoionization of the propylene oxide. Besides, for all investigated fragmentation channels the recorded KER distributions indicate that the excess of the used photon energy, with respect to the measured double ionization threshold energy of 28.3±0.1 eV, should be released as kinetic energy of ejected electrons. The bimodality of the KER for reaction (3) suggests the occurrence of two fragmentation mechanisms of the precursor molecular dication. Actually, the $CH_3^+$ detachment can leave the other ion in a variety of internal energy states. Even possible is a rearrangment of the molecular bonds due to atom migration. This is suggested by the observation of a bimodality in the total KER distribution for the $CH_3^+$ + $C_2H_3O^+$ product ions of reaction (3). Further experiments performed using isotopically labeled precursor molecules should clarify the relative importance of the concurrent alternative pathways. The perspective opens for planning measurements of the angular distributions of the final ions,[46] in order to investigate in greater detail the microscopic two-body dissociation mechanisms. Further, theoretical efforts will be done by our group, by using the same methodology already applied for different investigated systems,[42] in order to calculate the energy and structure of dissociation product ions to provide additional information on the dynamics of the charge separation reactions following the photoionization event. Interpretational techniques, as developed for photodissociation into neutral fragments on involved surfaces with their couplings and associated dynamical simulations, could be of interest for the bimodality of KER observed in the present case.

**ACKNOWLEDGMENTS**

Fernando Pirani acknowledges funding from MIUR, "Ministero dell'Istruzione, dell'Università e della Ricerca", PRIN 2015 (*STARS in the CAOS- Simulation Tools for Astrochemical Reactivity and Spectroscopy in the Cyberinfrastructurefor Astrochemical Organic Species*, 2015F59J3R). MIUR is also gratefully acknowledged for financial supporting through SIR 2014 "Scientific Independence for young Researchers" (RBSI14U3VF).

**Table I**

Products of the double ionization of propylene oxide, their abundance, averaged over the photon energy investigated and the relevant threshold energies.

| | dissociation ion products | average abundance (%) | threshold energy (eV) | $E_{max}$[a] kinetic energy of lighter ion (eV) | $E_{max}$[a] kinetic energy of heavier ion (eV) | $E_{max}$[a] total kinetic energy (eV) |
|---|---|---|---|---|---|---|
| 1 | $C_2H_4^+ + CH_2O^+$ | 66.70 | 28.3±0.1 | 2.5±0.2 | 2.3±0.2 | 4.9±0.3 |
| 2 | $CH_2^+ + C_2H_4O^+$ | 7.84 | 28.5±0.1 | 3.1±0.2 | 1.0±0.2 | 4.2±0.3 |
| 3 | $CH_3^+ + C_2H_3O^+$ | 5.00 | 29.0±0.3 | 4.5±0.2 | 1.5±0.2 | 6.1±0.3 |
| 4 | $O^+ + C_3H_6^+$ | 1.59 | 29.0±0.2 | 3.3±0.2 | 1.2±0.2 | 4.7±0.3 |
| 5 | $C_2H_3^+ + CH_3O^+$ | 18.70 | 29.2±0.1 | 2.8±0.2 | 2.4±0.2 | 5.3±0.3 |
| 6 | $OH^+ + C_3H_5^+$ | 0.17 | 32.1±0.3 | [b] | [b] | [b] |

[a] The value of the KER, $E_{max}$, at the maximum of the distributions.
[b] Too weak to extract the KER distribution.

**CAPTIONS TO THE FIGURES**

Fig. 1 - A scheme of the ion extraction and detection system used for the photoelectron-photoion-photoion coincidence measurements. In the Figure MCP detector stands for Multi Channel Plate detector, and the molecular beam source is a stainless steel needle effusive beam source with a nozzle having a 1.0 mm diameter.

Fig. 2 - A portion of the mass spectrum and ion–ion time of flight correlation of ions produced by single and double photoionization of propylene oxide at 37.0 eV. The coincidence plot is related to the main recorded fragmentation channels leading to $C_2H_4^+$ + $CH_2O^+$ and $C_2H_3^+$ + $CH_3O^+$ product ions. In this type of plot, which is typical of double photoionization experiments, the two time-of-flight values of a couple of ions produced in the same photoionization event define a point. In the Figure the dots region used for the evaluation of both the $C_2H_3^+/CH_3O^+$ and $C_2H_4^+/CH_2O^+$ kinetic-energy-released distributions is indicated by a white dashed oval.

Fig. 3 - (a) The relative cross section for the two body fragmentation channels accessible in the double photoionization of propylene oxide in the 18.0-37.0 eV photon energy range. The process leading to the $OH^+$ + $C_3H_5^+$, also observed, has not been reported in the graph, because too weak (see Table I). Lines interpolating points have been inserted only to better emphasize the observed trends. The arrow in the Figure indicates the microscopic path direction for the two body dissociation of the $(C_3H_6O)^{2+}$ dication towards the various ion pair products indicated on the right part of the Figure. (b) The threshold energy for the ion-ion dissociative processes following the double photoionization of propylene oxide. The process leading to the $OH^+$ + $C_3H_5^+$, also observed, has not been reported in the graph, because it is of too weak intensity (see Table I).

Fig. 4 - The KER distributions of the $CH_2^+ + C_2H_4O^+$ products formed in the double photoionization of propylene oxide, at different photon energy.

Fig. 5 - The total KER distributions of the $CH_3^+ + C_2H_3O^+$ products formed in the double photoionization of propylene oxide, at two different photon energies (35.0 and 37.0 eV). In the Figure the data are best fitted (full line) by a sum of two Gaussian functions (dashed line) indicating a bimodal behavior fingerprint of two alternative microscopic mechanisms.

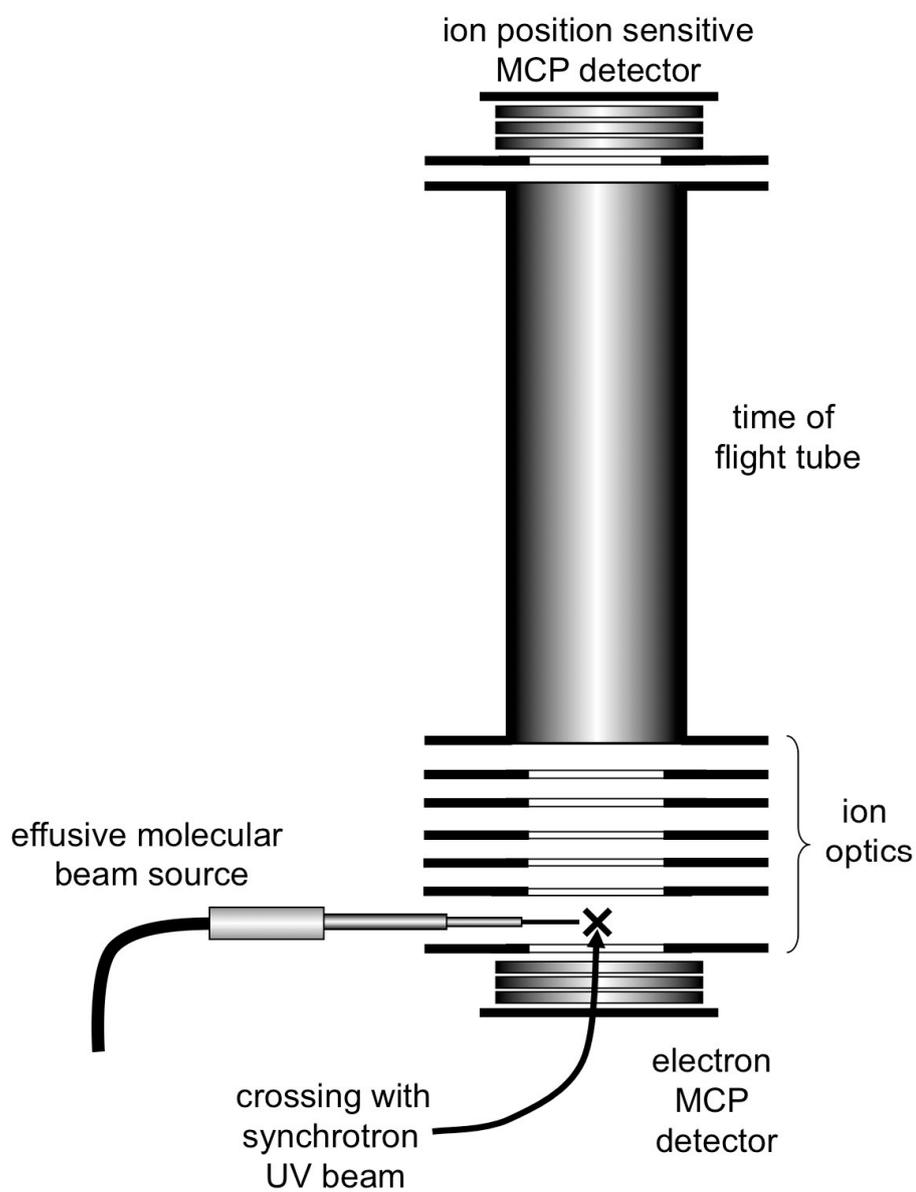

Fig. 1 - A scheme of the ion extraction and detection system used for the photoelectron-photoion-photoion coincidence measurements. In the Figure MCP detector stands for Multi Channel Plate detector, and the molecular beam source is a stainless steel needle effusive beam source with a nozzle having a 1.0 mm diameter.

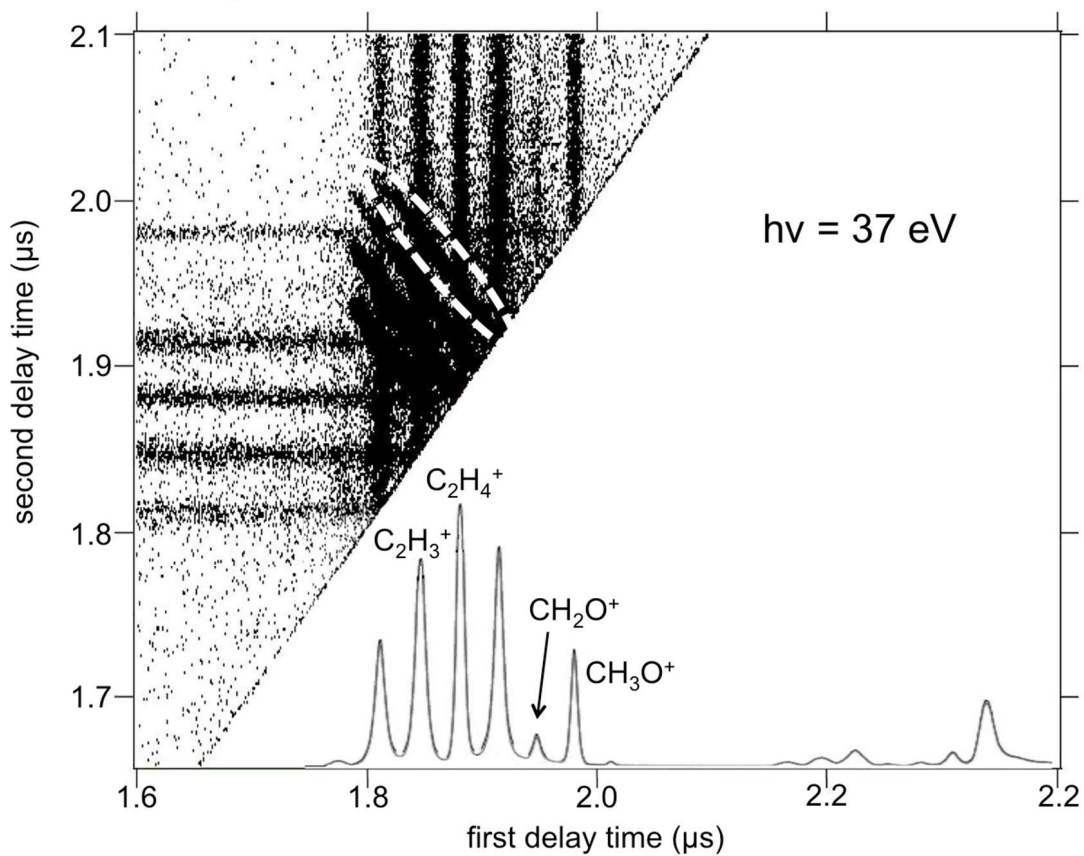

Fig. 2 - A portion of the mass spectrum and ion–ion time of flight correlation of ions produced by single and double photoionization of propylene oxide at 37.0 eV. The coincidence plot is related to the main recorded fragmentation channels leading to $C_2H_4^+$ + $CH_2O^+$ and $C_2H_3^+$ + $CH_3O^+$ product ions. In this type of plot, which is typical of double photoionization experiments, the two time-of-flight values of a couple of ions produced in the same photoionization event define a point. In the Figure the dots region used for the evaluation of both the $C_2H_3^+/CH_3O^+$ and $C_2H_4^+/CH_2O^+$ kinetic-energy-released distributions is indicated by a white dashed oval.

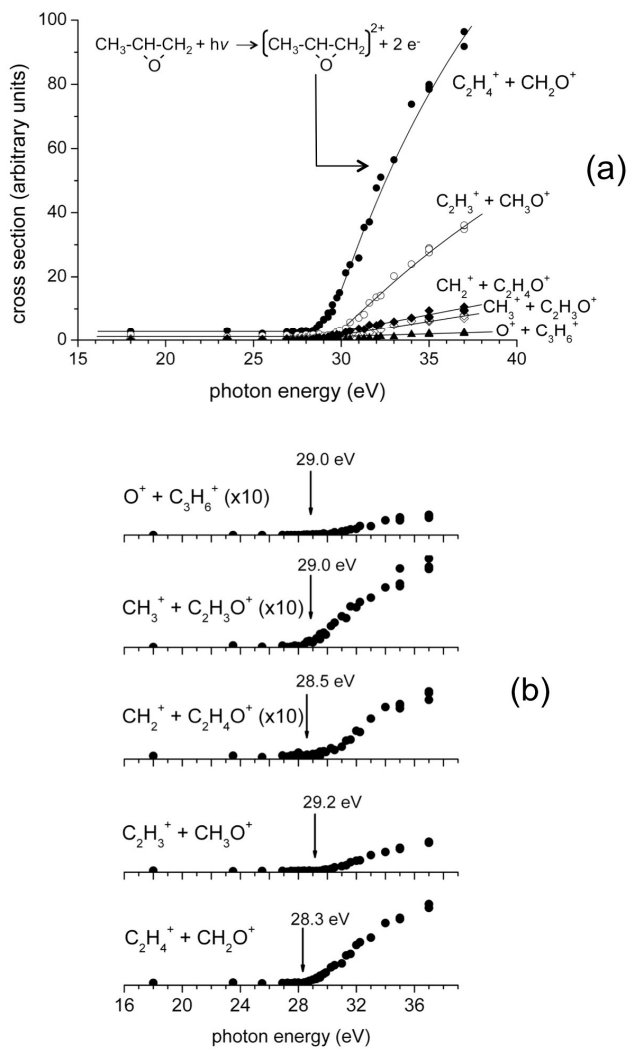

Fig. 3 - (a) The relative cross section for the two body fragmentation channels accessible in the double photoionization of propylene oxide in the 18.0-37.0 eV photon energy range. The process leading to the $OH^+ + C_3H_5^+$, also observed, has not been reported in the graph, because too weak (see Table I). Lines interpolating points have been inserted only to better emphasize the observed trends. The arrow in the Figure indicates the microscopic path direction for the two body dissociation of the $(C_3H_6O)^{2+}$ dication towards the various ion pair products indicated on the right part of the Figure. (b) The threshold energy for the ion-ion dissociative processes following the double photoionization of propylene oxide. The process leading to the $OH^+ + C_3H_5^+$, also observed, has not been reported in the graph, because it is of too weak intensity (see Table I).

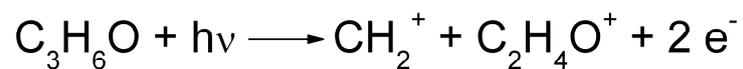

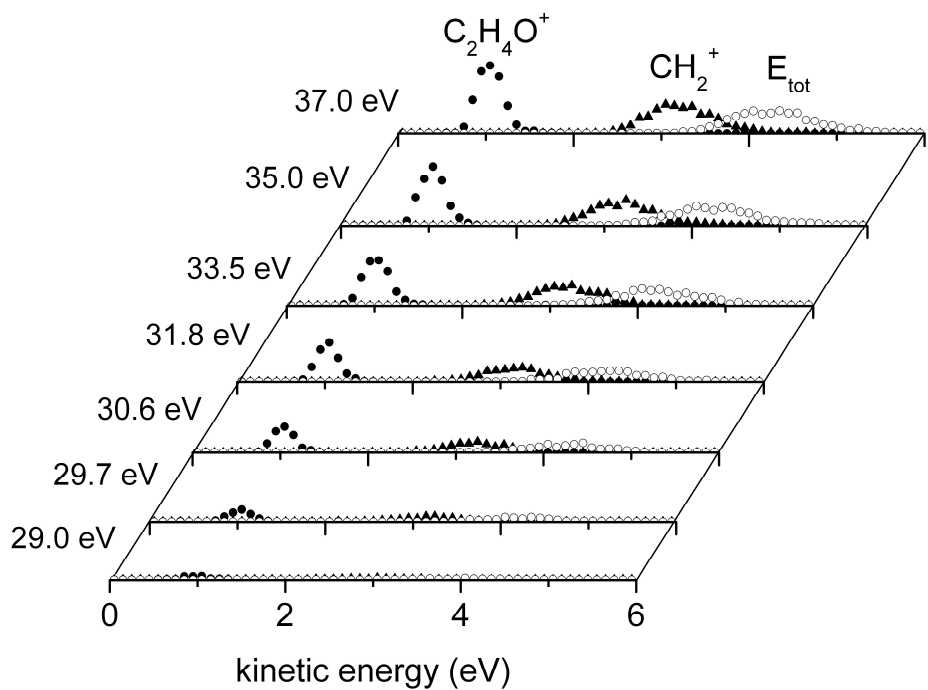

Fig. 4 - The KER distributions of the $CH_2^+ + C_2H_4O^+$ products formed in the double photoionization of propylene oxide, at different photon energy.

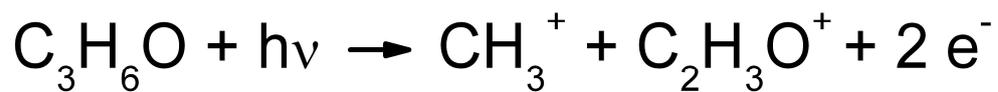

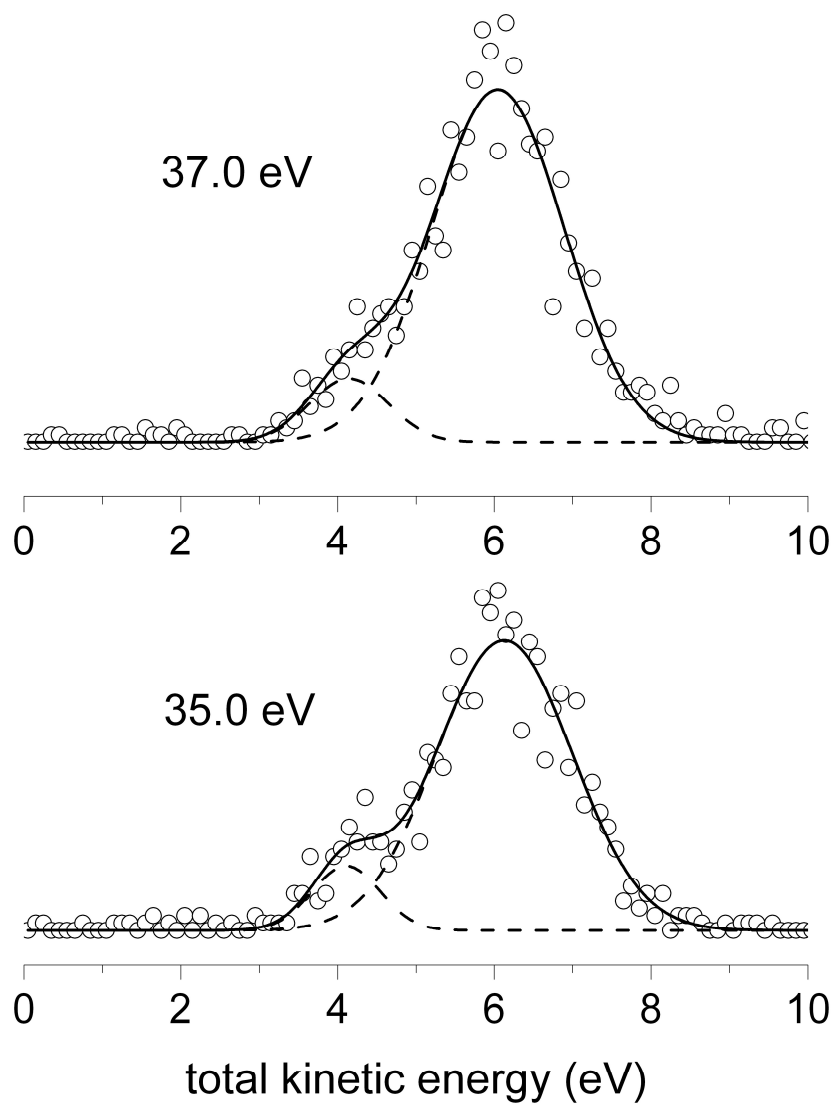

Fig. 5 - The total KER distributions of the $CH_3^+ + C_2H_3O^+$ products formed in the double photoionization of propylene oxide, at two different photon energies (35.0 and 37.0 eV). In the Figure the data are best fitted (full line) by a sum of two Gaussian functions (dashed line) indicating a bimodal behavior fingerprint of two alternative microscopic mechanisms.

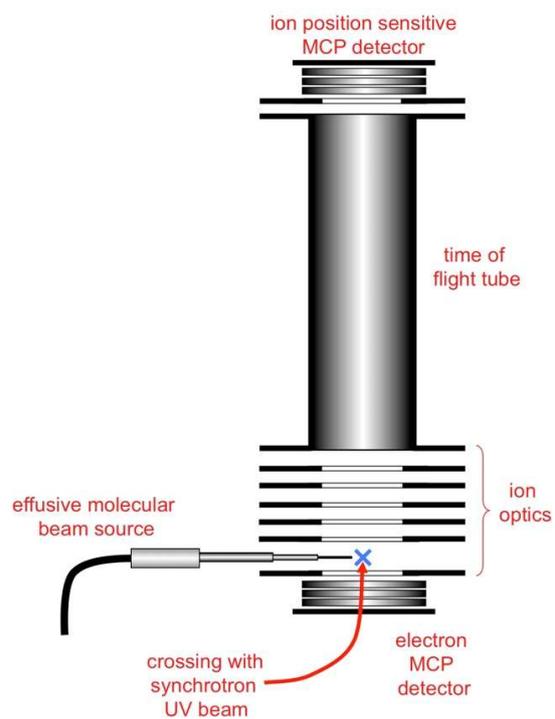

GRAPHICAL ABSTRACT